\documentclass[aps,prb,showpacs,twocolumn,superscriptaddress]{revtex4-1}

\usepackage{amsmath}
\usepackage{amssymb,amscd,bm,dsfont,wasysym,mathrsfs,latexsym,psfrag,accents,mathtools}
\usepackage{graphicx}
\usepackage{multirow}
\usepackage{dcolumn}
\usepackage{float}
\usepackage{color}
\usepackage{xcolor}
\usepackage{amsthm}

\newcolumntype{L}[1]{>{\raggedright\arraybackslash}p{#1}}
\newcolumntype{C}[1]{>{\centering\arraybackslash}p{#1}}
\newcolumntype{R}[1]{>{\raggedleft\arraybackslash}p{#1}}

			\newcommand{\e}[1]{\begin{align}{#1}\end{align}}	
		
		\newcommand{\f}[2]{\frac{#1}{#2}}
		\newcommand{\tf}[2]{\tfrac{#1}{#2}}
		
		\newcommand{\la}[1]{\label{#1}}

		\newcommand{\q}[1]{Eq.\ (\ref{#1})}
		\newcommand{\qq}[2]{Eqs.\ (\ref{#1}),(\ref{#2})}
		\newcommand{\qqq}[2]{Eqs.\ (\ref{#1})-(\ref{#2})}
		
		\newcommand{\fig}[1]{Fig.\ \ref{#1}}

		\newcommand{\eq}{=&\;}

		\newcommand{\R}{\mathbb{R}}

	\newcommand{\eikr}{e^{i\bk \cdot \br}}

\newcommand{\nabk}{\nabla_{\boldsymbol{k}}}

\newcommand{\bt}{$BT_{\sma{\perp}}$}
\newcommand{\bts}{$BT_{\sma{\perp}}$ }

\newcommand{\var}{\varepsilon}

\newcommand\as{\;\;\;\;}

\newcommand{\hbr}{\hat{\br}}

\newcommand{\bk}{\boldsymbol{k}}
\newcommand{\bkp}{\boldsymbol{k}^{{\sma{\perp}}}}

\newcommand{\bp}{\boldsymbol{p}}

\newcommand{\br}{\boldsymbol{r}}
\newcommand{\bs}{\boldsymbol{s}}

\newcommand{\bv}{\boldsymbol{v}}

\newcommand{\bB}{\boldsymbol{B}}

\newcommand{\bK}{\boldsymbol{K}}

\newcommand{\bze}{\boldsymbol{0}}

\newcommand{\bPi}{\boldsymbol{\Pi}}

\newcommand{\frako}{\mathfrak{o}}

\newcommand{\orb}{\boldsymbol{\mathfrak{A}}}

\newcommand\rot{\mathfrak{c}}
\newcommand\scr{\mathfrak{s}}

\newcommand{\sx}{\sigma_{\sma{1}}}

\newcommand{\sz}{\sigma_{\sma{3}}}

\newcommand{\calh}{{\cal H}}

\newcommand{\calt}{{\cal T}}

\newcommand{\mo}{\text{-}1}

\newcommand{\minus}{\text{-}}

\newcommand{\braket}[2]{\big\langle #1 \big| #2 \big\rangle}

\newcommand{\bra}[1]{\big\langle#1\big|}
\newcommand{\ket}[1]{\big|#1\big\rangle}

\newcommand{\lin}{\notag \\}

\newcommand{\bpm}{\begin{pmatrix}}
\newcommand{\epm}{\end{pmatrix}}

\newcommand{\bal}{\begin{align}}

\newcommand{\sma}[1]{\scriptscriptstyle{#1}}

\newcommand{\Z}{\mathbb{Z}}

\begin{document}

\title{Modern theory of magnetic breakdown}
 \author{A. Alexandradinata} \affiliation{Department of Physics, Yale University, New Haven, Connecticut 06520, USA}  
  \author{Leonid Glazman} \affiliation{Department of Physics, Yale University, New Haven, Connecticut 06520, USA}  
  

\begin{abstract}
The modern semiclassical theory of a Bloch electron in a magnetic field  encompasses  the orbital magnetization and geometric phase. Beyond this semiclassical theory lies the quantum description of field-induced tunneling between semiclassical orbits, known as magnetic breakdown. Here, we synthesize the modern semiclassical notions with quantum tunneling -- into a single Bohr-Sommerfeld quantization rule that is predictive of magnetic energy levels. This rule  is applicable to a host of topological solids with \emph{unremovable} geometric phase, that also \emph{unavoidably} undergo breakdown. A notion of topological invariants is formulated  that nonperturbatively encode tunneling, and is measurable in the de-Haas-van-Alphen effect. Case studies are discussed for topological metals near a metal-insulator transition and over-tilted Weyl fermions. 
\end{abstract}
\date{\today}


\maketitle

The semiclassical Peierls-Onsager-Lifshitz theory\cite{peierls_substitution,onsager,luttinger_peierlssub,lifshitz_kosevich} of a Bloch electron in a magnetic field has been extended\cite{kohn_effham,rothI,blount_effham} to incorporate two modern notions: a wavepacket orbiting in quasimomentum $(\bk)$ space acquires a geometric Berry phase ($\phi_{\sma{B}}$),\cite{berry1984,mikitik_berryinmetal}  as well as a second phase ($\phi_{\sma{R}}$) originating from the orbital magnetic moment of a wavepacket about its center of mass.\cite{chang_niu_hyperorbit,wilkinson_semiclassical_harper} Both $\phi_{\sma{B}}$ and $\phi_{\sma{R}}$ are evaluated on semiclassical orbits which are uniquely determined by Hamilton's equation. If the quasimomentum separation between two neighboring orbits is of the order of the inverse magnetic length, field-induced quantum tunneling (known as magnetic breakdown)\cite{cohen_falicov_breakdown,blount_effham,azbel_quasiclassical,pippard1,pippard2,chambers_breakdown,slutskin} invalidates a unique semiclassical trajectory. 
 
Can the modern semiclassical notions of geometric phase and orbital moment be combined with the quantum phenomenon of breakdown? A unified theory would describe a host of solids which have emerged in the recent intercourse between band theory and topology. These solids are characterized by geometric phase which is unremovable owing to symmetry; the robust intersection of orbits simultaneously guarantees breakdown.

 We propose that the magnetic energy levels in these solids are determined by Bohr-Sommerfeld quantization rules that unify tunneling, geometric phase and the orbital moment -- these rules generalize  the Onsager-Lifshitz-Roth quantization rules\cite{onsager,lifshitz_kosevich,rothI} for transport within a single band, and provide an algebraic method to calculate Landau-level spectra without recourse\cite{Serbyn_LandauofTCI,obrien_breakdown,koshino_figureofeight} to large-scale, numerical diagonalization. These rules are also predictive of  de-Haas-van-Alphen\cite{dHvA,SdH} (dHvA) peaks, as well as of fixed-bias peaks of the differential conductance in scanning-tunneling microscopy (STM).\cite{Sangjun_Cd3As2,Ilija_SnTe}


 While oscillatory patterns in the dHvA\cite{dHvA,SdH} measurement underlie the `fermiological'\cite{shoenberg} construction of Fermi surfaces,\cite{ashcroft_mermin,champel_mineev} such oscillations are generically disrupted by tunneling in low-symmetry solids.\cite{kaganov_coherentmagneticbreakdown} Here, we demonstrate how multi-harmonic oscillations may nevertheless persist in high-symmetry solids whose orbits intersect at a saddlepoint. Furthermore, the phase offset of each harmonic is a topological invariant that non-perturbatively encodes quantum tunneling in magnetotransport, as well as sharply distinguishes metals with differing Berry phases on their Fermi surface.  

Our last case study describes tunneling which occurs at the intersection of a hole and electron pocket, as exemplified by an over-tilted Weyl point;\cite{isobe_nagaosa_IIDirac,Bergholtz_typeII,WTe2Weyl,LMAA} the corresponding magnetic energy levels  were first studied numerically in Ref.\ \onlinecite{obrien_breakdown}. Here, we present the first Berry-phase-corrected quantization rule which is valid at any tunneling strength, and compare our algebraically-derived Landau-level spectra to their\cite{obrien_breakdown} numerically-exact spectra.

\begin{figure}[ht]
\centering
\includegraphics[width=8.3 cm]{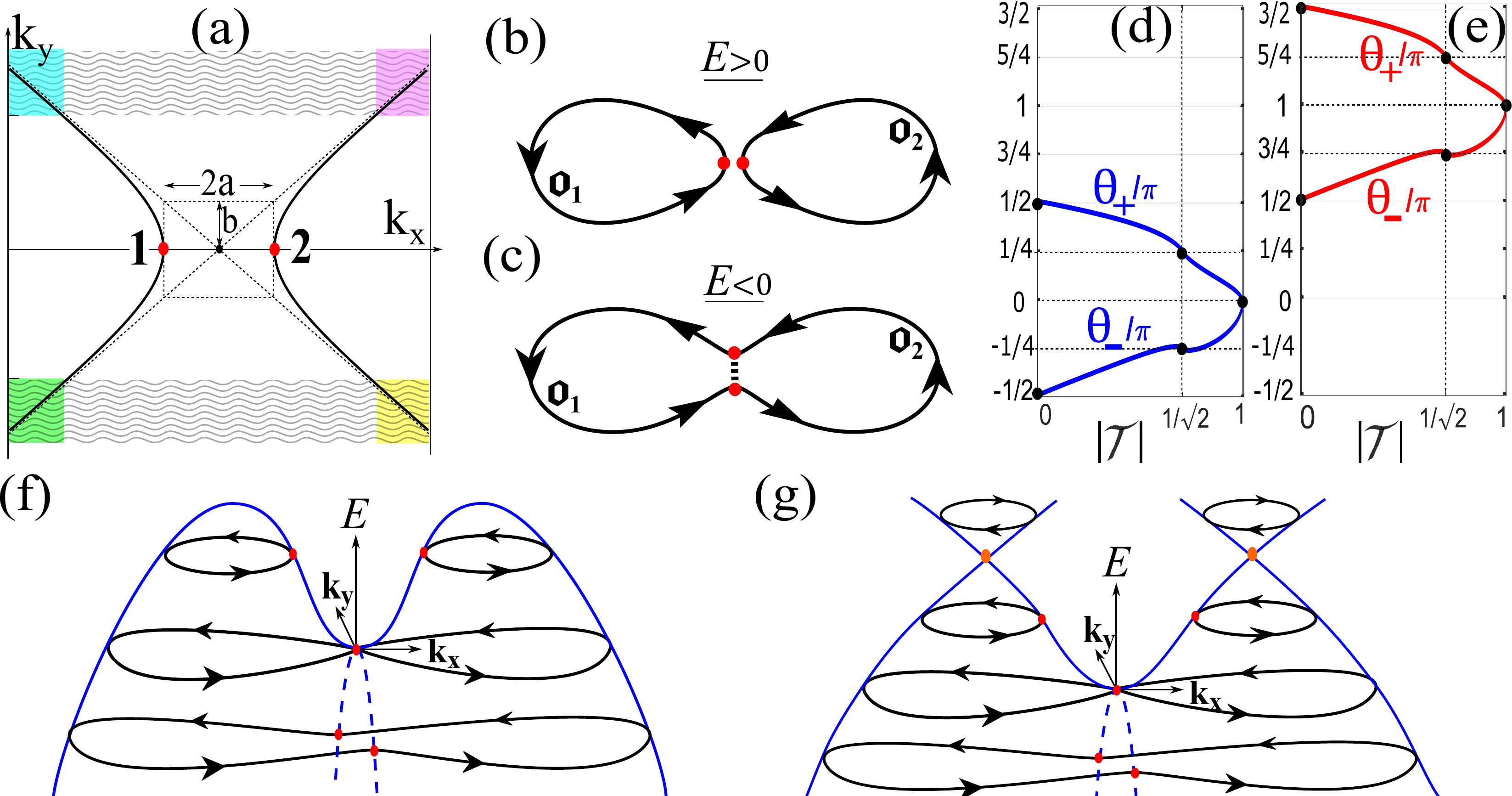}
\caption{(a) illustrates a region in $\bkp$-space with significant tunneling. (b-c) Constant-energy band contours of two distinct orbits (at positive energy) that merge into one (at negative energy). Black arrows indicate the orientation determined by Hamilton's equation. (d-e) Plot  of  $\theta_{\eta}$ vs $|\calt|{=}(1{+}e^{\minus 2\pi{\mu}})^{\minus\sma{1/2}}$ for the  conventional metal (blue line) and the topological metal (red). (f-g) Band dispersion of a conventional metal (left), and a topological metal (right) with  two Dirac points. }\label{fig:intraband}
\end{figure}


We shall assume throughout this letter that the field is oriented in $\vec{z}$, such that orbits are contours of the band dispersion at fixed energy $E$ and $k_z$. In the $\bkp{:=}(k_x,k_y)$-neighborhood where tunneling occurs, two orbits approach each other as two arms of a hyperbola illustrated in \fig{fig:intraband}(a); tunneling is significant if the area $(4ab)$ of the rectangle inscribed between the  arms is comparable or smaller than $1/l^2$, with $l{:}{=}(\hbar c/e|\bB|)^{\sma{1/2}}$ the magnetic length. The orientation of the approaching orbits, as determined by Hamilton's equation $\hbar \dot{\bk}{=}{-}|e| \bv {\times} \bB/\hbar c$, distinguishes between two qualitatively distinct types of breakdown: (a) if both arms carry the same sense of circulation, tunneling occurs between contours belonging to the same band. Case (b) for which both arms are oppositely oriented will be discussed in the second half of the letter. The former case is known as intraband breakdown, which occurs wherever band contours  change discontinuously as a function of energy; the nucleus of this Lifshitz transition is a saddlepoint which disperses as $\var_{\bk}{=}k_{\sma{x}}^{\sma{2}}/2m_1{-}k_{\sma{y}}^{\sma{2}}/2m_2$. The vanishing band velocity at $\bkp{=}\bze$ implies that a hypothetical wavepacket satisfying Hamilton's equation never reaches the saddlepoint in finite time.\cite{kosevich_topologyreview} The probability of vertical transmission (between ${\nearrow}$-incoming and $\nwarrow$-outgoing trajectories) equals $|\calt|^2{:=}(1{+}e^{\minus 2\pi{\mu}})^{\minus\sma{1}}$,\cite{kemble} with ${\mu}{:=}\sqrt{m_1m_2}El^2$ geometrically interpreted as $abl^2/2$, and $E$ measured from the saddlepoint.


 

The conceptually-simplest realization of intraband breakdown occurs for two orbits (at $E{>}0$) that merge into a single orbit (at $E{<}0$), as illustrated in \fig{fig:intraband}(b-c).  This merger has at least two topologically-distinct realizations: (ai) a conventional metal whose band dispersion has two nearby maxima [\fig{fig:intraband}(f)], and (aii) a topological metal near a metal-insulator transition, where two Weyl points (two-band touching points with conical dispersions)\cite{NIELSEN1981A,wan2010} with opposite chirality are near annihilation [\fig{fig:intraband}(g)].  Our comparative study of (ai-aii) illustrates how their difference in Berry phase manifests in the magnetic energy levels, which are determined in both cases by the following quantization rule:
\e{\cos\left[\tf{\Omega_1+\Omega_2}{2}\big|_{E,l^2}+\varphi({\mu}) \right] = |\calt({\mu})|\cos\left[\tf{\Omega_1-\Omega_2}{2}\big|_{E,l^2} \right]. \la{quantizationnecking}}
$\calt{:}{=}(1{+}e^{\minus 2\pi{\mu}})^{\minus\sma{1/2}}e^{i\varphi}$ is the aforementioned amplitude for vertical transmission, with $\varphi{=}\arg[\Gamma(1/2{-}i\mu)]{+}\mu \log |\mu|{-}\mu$ involving the Gamma function.\footnote{$|\varphi|$ is bounded by $0.05\pi$} $\Omega_j{:}{=}\Omega[\frako_j]$ is the semiclassical phase acquired by a wavepacket in traversing ${\frako}_{j}$, which is a closed Feynman trajectory illustrated in \fig{fig:intraband}(b-c). For $E{>}0$, $\frako_1$ is simply the left orbit in \fig{fig:intraband}(b); for $E{<}0$, $\frako_1$ combines the left half of the orbit with a tunneling trajectory [dashed line in \fig{fig:intraband}(c)] through the semiclassically-forbidden region. 
\e{\Omega[\frako_j(E),l^2] = l^2S[\frako_j(E)\big]+\phi_{\sma{M}}+\lambda\big[\frako_j(E)\big], \la{definenujberryonly}}
which includes (i) a dynamical phase that is proportional to the $\bkp$-space area $S$ bounded by ${\frako}_j$, with  $S$ being positive (resp.\ negative) for a clockwise-oriented (resp.\ anticlockwise) orbit. Here, $\{S[\frako_j]\}$ carry the same sign. 


The remaining contributions to $\Omega_j$ are subleading in powers of $|\bB|$: (ii) the Maslov phase ($\phi_{\sma{M}}$) equals $\pi$ for trajectories that are deformable to a circle,\cite{keller1958} and (iii) a further correction $(\lambda)$ encodes the aforementioned geometric phase and orbital moment, as well as the well-known Zeeman coupling. Whether the orbital moment contributes to $\lambda$ depends on the crystalline symmetries of the spin-orbit-coupled solid, as well as the field alignment with respect to certain crystallographic axes. Let us first consider a time-reversal-invariant  but non-centrosymmetric metal, with a two-fold rotational   axis  that is parallel to the applied field ($\vec{z}$) -- these symmetries stabilize Weyl points in the rotationally-invariant two-torus (denoted \bt).\cite{WTe2Weyl} Then, $\lambda_j{:}{=}\lambda[\frako_j]$ equals the geometric phase ($\phi_{\sma{B}}$), which is the line integral over ${\frako}_j$ of the Berry one-form\cite{berry1984} $i\braket{u_{1\bk}}{\nabk u_{1\bk}}{\cdot} d\bk$. Here, $\eikr u_{1\bk}$ is the Bloch function of the  low-energy band drawn in \fig{fig:intraband}(f-g); one may verify that redefining $u_{1\bk}$ by a $\bk$-dependent phase  may add to $\phi_{\sma{B}}$ an integer multiple of $2\pi$, but does not affect the quantization condition in \q{quantizationnecking}. The composition of time-reversal and two-fold rotation is a symmetry (denoted $T\rot_{2z}$) that essentially makes wavefunctions real at each $\bkp{\in}$\bt, hence $e^{i\phi_{\sma{B}}}{\in}\R$,\footnote{Each orbit falls into symmetry class [I,$s{=}0$] in the ten-fold classification of Ref.\ \onlinecite{topofermiology}, where  the resultant symmetry constraints on the Berry phase, orbital moment and Zeeman effect are explained in full generality.}  with $\phi_{\sma{B}}{=}0$ and $\pi$ for the conventional and topological metal respectively. Moreover, since the $z$-component of angular momentum flips under $T\rot_{2z}$, the orbital moment lies parallel to \bts and does not contribute to $\lambda$.\cite{topofermiology} Applying the same argument to the expectation value [$\bs(\bk)$] of the spin operator with respect to $u_{1\bk}$, we derive that the Zeeman coupling, being proportional to $s_z$, is also trivial. 

To observe the effects of the orbital moment and Zeeman coupling, we consider a different symmetry class of solids with a mirror symmetry ($x{\rightarrow}{-}x$) that relates the two maxima in (ai) and the two Weyl points in (aii); this symmetry allows the orbital moment/$\bs$ to tilt out of \bts at  $\bkp$ which are not reflection-invariant. Then $\lambda{=} \phi_{\sma{B}}{+}\phi_{\sma{R}}{+}\phi_{\sma{Z}}$,\footnote{The present form of $\Omega_j$ presupposes a reflection symmetry; the most general form of $\Omega_j$ is clarified in a separate publication\cite{AALG}.} with $\phi_{\sma{R}}$ defined as the line integral of the orbital-moment one-form:\cite{topofermiology}
\e{ \orb \cdot d\bk =  i\sum_{l\neq 1} {\braket{u_{1\bk}}{\nabk u_{l\bk}}\Pi^y_{l1}} {dk_x}/{2{v}_y}+ (x \leftrightarrow y). \la{definerothoneform}}
Here, $\bPi(\bk)_{ln}{:}{=}i\bra{u_{l\bk}}e^{-i\bk \cdot \hbr}[\hat{H}_0,\hbr]e^{i\bk \cdot \hbr}\ket{u_{n\bk}}/\hbar$ are matrix elements of the  velocity operator, $\bv{:}{=}\bPi_{11}$, and $\hat{H}_0$ is the single-particle, translation-invariant Hamiltonian, and $\hbr$ the position operator.  ${\sum}_{\sma{l\neq 1}}$ denotes a sum over all bands excluding $u_{1\bk}$.\footnote{These other bands are not illustrated in \fig{fig:intraband}(f) but exist as a matter of principle; in \fig{fig:intraband}(g), only one other band is illustrated.}   Finally, $\lambda$ is contributed by the Zeeman phase ($\phi_{\sma{Z}}$), which is the line integral of  $g_0 s_z(\bk) {|d\bk|}/2m(v^{\sma{2}}_x{+}v^{\sma{2}}_y)^{\sma{1/2}}$, with $g_0{\approx}2$ the free-electron g-factor, and $m$ the free-electron mass. If the orbital moment/$\bs$ tilts toward ${+}\vec{z}$ at a wavevector $\bkp {\in}\frako_1$, the tilt occurs toward ${-}\vec{z}$ in the reflection-mapped wavevector lying in $\frako_2$, hence $\lambda_1{=}{-}\lambda_2$ modulo $2\pi$.\footnote{Both orbits collectively fall into symmetry class [II-B,$s{=}0,u{=}1$] in Ref.\ \onlinecite{topofermiology}.}



The quantization rule [\q{quantizationnecking}] has been derived by Azbel\cite{azbel_quasiclassical} in the Peierls-Onsager approximation,\cite{peierls_substitution,onsager,luttinger_peierlssub} which effectively dispenses with the $\lambda$-correction to $\Omega$.   By accounting for the subleading-in-$|\bB|$ correction\cite{kohn_effham,rothI,blount_effham} to the effective Hamiltonian of a Bloch electron in a magnetic field, we have derived an improved connection formula\cite{AALG} for the WKB wavefunction,\cite{zilberman_wkb, fischbeck_review} which is valid only in the semiclassical $\bkp$-regions [indicated by wavy lines in \fig{fig:intraband}(a)] where tunneling is negligible. Continuity of the connected WKB wavefunction imposes the improved quantization rule in \q{quantizationnecking}, inclusive of the $\lambda$-correction.

When \q{quantizationnecking} is viewed at fixed field, the discrete energetic solutions   correspond  to Landau levels. When  viewed at constant Fermi energy ($E_F$), the discrete solutions correspond to values of $l^2$ where Landau levels successively become equal to the Fermi energy, leading to peaks in a dHvA or fixed-bias STM measurement; such discrete $l^2$ will henceforth be referred to as dHvA levels. Some intuition about the Landau/dHvA levels may be gained in the  semiclassical limit: $\mu{\rightarrow} \infty$, where $\calt {\rightarrow} 1$, and \q{quantizationnecking} simplifies to independent quantization rules for two uncoupled orbits $\frako_j$ illustrated in \fig{fig:intraband}(b): $\Omega_j/2\pi{\in}\Z$.  The Landau spectrum splits into two sets labelled by $j$, where adjacent spacings within each set are locally periodic as $E_{j,n+1}{-}E_{j,n}{=}{2\pi }/[{l^2(\partial S_j/\partial E)}$] with the right-hand-side evaluated at $E_{j,n}$, $n{\in}\Z$ and $S_j{:}{=}S[\frako_j]$. Analogously, the dHvA levels split into two sets, where adjacent levels in each set are periodic as $l^2_{j,n+1}{-}l^2_{j,n}{=}2\pi/S_j(E_F)$. This (local) periodicity also characterizes  the opposite semiclassical limit $\mu{\rightarrow}{-}\infty$, where both $\calt$ and $\phi {\rightarrow} 0$, and we obtain a single quantization rule for the combined orbit ${\frako}_1{+}{\frako}_2$ illustrated in \fig{fig:intraband}(c).  Let us describe the case of general $\mu$ in symmery classes where the two orbits are not mutually constrained (this includes the $T\rot_{2z}$ class): the two incommensurate harmonics $(\Omega_1{\pm}\Omega_2)/2$ in \q{quantizationnecking} then competitively produce a Landau/dHvA spectrum that is not (locally) periodic but retains a long-ranged correlation; such spectra have been called quasirandom.\cite{kaganov_coherentmagneticbreakdown} 


In contrast, the mirror symmetry in the second class of solids enforces $S[\frako_1]{=}S[\frako_2]{:}{=}S$ at all energies, and this demonstrably  allows for locally-periodic spectra. The mirror-symmetric quantization condition  is solved by two sets of Landau/dHvA levels distinguished by an index $\eta{\in}{\pm}$: $l^2|S(E)|{=} 2\pi n{+}\phi_{\sma{M}} {+} \theta_{\eta}$, with
\e{ \theta_{\eta}(E,l^2):=  \varphi({\mu})+  \cos_{\eta}^{\minus 1}\big(\;|\calt(\mu)|\cos\big(\lambda_1\big)\;\big),\la{principalquanti}}
defined as a phase: $\theta {\sim}\theta{+}2\pi$, and $\cos^{\minus \sma{1}}_{\sma{\eta}}(\cdot)$ denotes the principal value lying in the interval $[0,\pi]$ for $\eta{=}{+}$, and in $[{-}\pi,0]$ for $\eta{=}{-}$. For $\mu{\rightarrow} \infty$, $\theta_{\pm}{\rightarrow}{\pm}\lambda_1$ implies a symmetric splitting of Landau levels; as $\mu{\rightarrow} {-}\infty$, $\theta_{\pm}{\rightarrow}{\pm}\pi/2$ implies that this symmetric splitting is exactly $\pi$, and both sets of Landau levels (distinguished by $\eta$) may be viewed as a single set of Landau levels with an emergent local period $2\pi/[l^2\partial (2S)/\partial E]$ -- this corresponds to a combined orbit that is intersected by a reflection-invariant line; $S[\frako_1{+}\frako_2]{=}2S$, and $\lambda[\frako_1{+}\frako_2]{=}0$.\footnote{This orbit falls into class [II-A,$u{=}1$,$s{=}0$] in Ref.\ \onlinecite{topofermiology}}  To observe locally-periodic dHvA levels at the Fermi energy, it is necessary that $\theta$ varies slowly on the scale of the dHvA period $2\pi/S(E_F)$. Indeed, the typical scale of variation  for $|\calt(\mu)|$ and $\varphi(\mu)$ is $\Delta \mu {\sim} 1$, which implies a dHvA scale $\Delta l^2 {\sim}  1/\sqrt{m_1m_2}E_F$ from the definition of $\mu$; $\Delta l^2/(2\pi/S(E_F))$ is therefore negligible for small enough $|E_F|$ or large enough $S(E_F)$. Presuming these conditions, $\theta_{\sma{\eta}}$ is extractable from the phase offset ($\gamma_{\sma{\eta}}{=}\theta_{\sma{\eta}}{+}\phi_{\sma{M}}{+}\phi_{\sma{LK}}$) of the $\eta$'th harmonic in the dHvA spectrum; the additional Lifshitz-Kosevich correction equals $\pi/4$ (resp.\ ${-}\pi/4$) for a minimal (resp.\ maximal) orbit in 3D solids.\cite{lifshitz_kosevich} \q{principalquanti} represents one key result for intraband breakdown -- that the dHvA phase offset nonlinearly depends on both the tunneling parameter $\calt$, as well as the semiclassical phase corrections: $\phi_{\sma{R}},\phi_{\sma{B}},\phi_{\sma{Z}}$.


To conclude our discussion of intraband breakdown, we propose a symmetry class where $\theta$ depends on a universal function of $\mu$, with an additive Berry-phase correction that is insensitive to symmetric deformations of the metal. In addition to the mirror symmetry presupposed in \q{principalquanti}, we further impose $T\rot_{2z}$ symmetry so that $e^{i\lambda}{=}e^{i\phi_{\sma{B}}}{=}1$ for the conventional metal, and ${-}1$ for the topological metal; this is, incidentally, the symmetry class of TaAs, which is known to have four mirror-related pairs\footnote{The orbits in TaAs may not resemble the ones illustrated in \fig{fig:intraband}(g), and a different quantization condition may be appropriate for TaAs\cite{AALG}.}  of Weyl points  in the rotational-invariant Brilloin torus.\cite{TaAs,ChinadiscoversTaAs,Princeton_discovers_TaAs} \q{principalquanti} thus simplifies to $\theta_{\eta}{=}\varphi(\mu){+}  \cos_{\eta}^{\minus 1}|\calt(\mu)| {+}\phi_{\sma{B}}$, which are plotted against $|\calt(\mu)|$ in \fig{fig:intraband}(d-e), for both topological (red line) and conventional (blue) metals. As $\mu$ is varied over $\R$, $\theta_{\sma{\eta}}$ robustly covers the interval $[\pi/2,3\pi/2]$ in the former case, and  $[-\pi/2,\pi/2]$ in the latter; the exact $\pi$ offset originated from the difference in Berry phase. In both cases, $\theta_{\sma{+}}{=}\theta_{\sma{-}}$ for $\mu{\rightarrow} \infty$ implies a two-fold-degeneracy in the Landau levels, which did not arise in the $T\rot_{2z}$-asymmetric case. We therefore associate the robust covering of a $\pi$-interval (in either case) to a Lifshitz transition in solids with $T\rot_{2z}$ and mirror symmetries. 	This may be viewed in a unifying analogy with field-free topological insulators, where the Berry phase covers $2\pi$\cite{Cohomological,Maryam2014,z2pack,fu_trspolarization,yu2011,soluyanov2011,AA2014} (or rational fractions thereof)\cite{berryphaseTCI,JHAA} as a function of a crystal wavevector. In comparison, $\theta_{\sma{\eta}}$ includes not just the Berry phase, but also  nonperturbatively encodes tunneling through its dependence on $\calt$.  Being robust against symmetry-preserving deformations of the metal, the $\pi$-covering of $\theta_{\sma{\eta}}$ may be viewed as a topological invariant in quantum magnetotransport.

\begin{figure}[ht]
\centering
\includegraphics[width=8.6 cm]{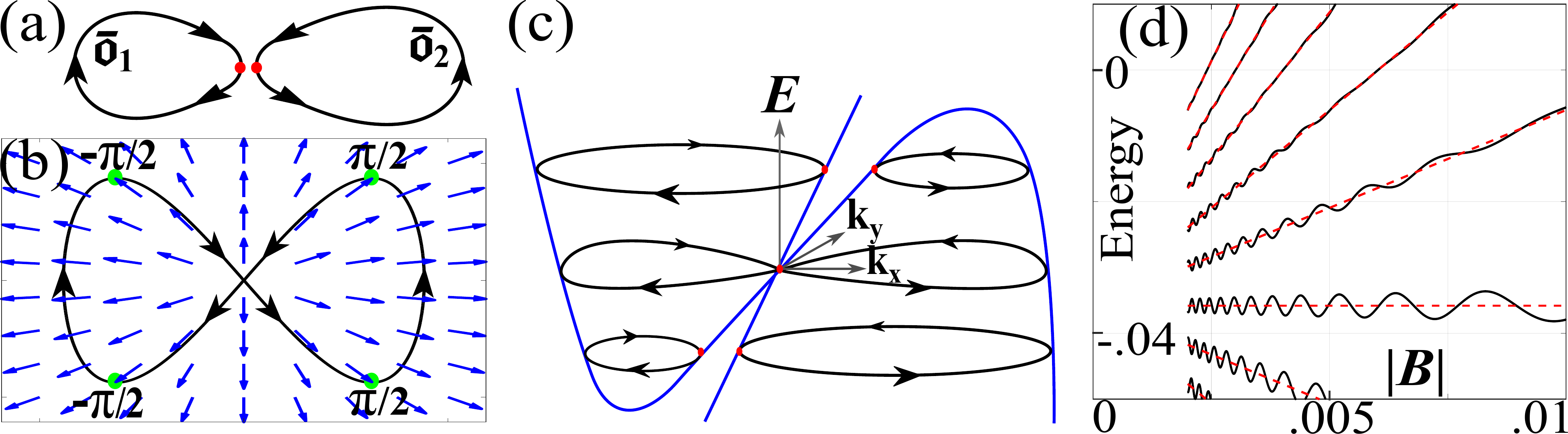}
\caption{ An over-tilted Weyl point that is not linked by tunneling to other Weyl points. (a) Band contours at fixed nonzero energy. (b) Zero-energy band contour of $H_{\sma{II}}{-}|t|(1{-}\sz)k_x^3{:}{=} d_1(\bkp)\sx{+}d_3(\bkp)\sz$; the pseudo-spin texture  is illustrated by blue arrows, with $d_1$ (resp.\ $d_3$) the vertical (resp.\ horizontal) component of each arrow.  (c) Band dispersion. (d) Dispersion of Landau levels for the tight-binding model in Ref.\ \onlinecite{obrien_breakdown}; we follow their units for energy and $|\bB|$.}\label{fig:interband}
\end{figure}

Let us next discuss case (b) where both hyperbolic arms are oppositely oriented, and the approaching orbits belong to distinct bands -- they touch at the intersection of hole-like and electron-like pockets, which is exemplified by an over-tilted Weyl fermion.\cite{isobe_nagaosa_IIDirac,Bergholtz_typeII,WTe2Weyl,LMAA} This touching point is modelled by the linearized Hamiltonian: $H_{\sma{II}}(\bkp){=} (u{+}v\sz)k_x{+}w\sx k_y$, with $|u|{>}|v|$ and $\sigma_j$ Pauli matrices spanning the two-band subspace. Interband tunneling occurs with the Landau-Zener probability $e^{\minus 2\pi\bar{\mu}}$,\cite{blount_effham} with $\bar{\mu}{=}{(vEl)^{\sma{2}}}/2|w|(u^{\sma{2}}-v^{\sma{2}})^{\sma{3/2}}$, and $E$ measured from the degeneracy;  in particular, this probability is unity at zero energy; in comparison, $|\calt|^2{=}1/2$ at the saddlepoint.\cite{kemble}  In both intra- and interband breakdown, the respective dimensionless parameters $|{\mu}|$ and $\bar{\mu}$ have the same geometrical interpretation as $abl^2/2$; however, $a$ and $b$ are distinct functions of $E$ and $\bk\cdot \bp$ parameters: $u,v,w$ in $H_{\sma{II}}$, and $m_1,m_2$ for the saddlepoint. 

The simplest scenario\cite{obrien_breakdown} of two orbits $\{\bar{\frako}_{\sma{j}}\}_{\sma{j=1}}^{\sma{2}}$ linked by interband breakdown describes an over-tilted Weyl fermion modelled by $H(\bkp){=}H_{\sma{II}}{-}|t|(1{-}\sz)k_x^3$; such fermions were predicted to arise in WTe$_2$,\cite{WTe2Weyl} whose symmetry class ($T\rot_{2z}$) we adopt in the following discussion.\footnote{WTe$_2$ is symmetric under a screw ($\scr$), which is composed of a two-fold rotation ($\rot_{2z}$) and a half lattice translation in $\vec{z}$. The group algebra involving $\scr$ is isomorphic to one involving $\rot_{2z}$, if we restrict ourselves to the $k_z=0$ plane.} The corresponding constant-energy band contours are illustrated in \fig{fig:interband}(a-c), and  the quantization condition is
\e{\cos\big[\tf{{\Omega}_{\bar{1}}+{\Omega}_{\bar{2}}}{2}\big|_{E,l^2}\big] = \tau(\bar{\mu})\cos\big[\tf{{\Omega}_{\bar{1}}-{\Omega}_{\bar{2}}}{2}\big|_{E,l^2} +\bar{\varphi}(\bar{\mu})\big], \la{quantizationIIdirac}}
where $\tau e^{i\bar{\varphi}}$ (with $\tau{:}{=}\sqrt{1{-}e^{\minus 2\pi\bar{\mu}}}$) is the amplitude of intraband transmission between $\searrow$-incoming and $\swarrow$-outgoing trajectories, and $\bar{\varphi}{=}\bar{\mu} {-}\bar{\mu} \ln \bar{\mu} {+}\text{arg}[ \Gamma(i\bar{\mu})] {+}\pi/4$. ${\Omega}_{\bar{j}}$ is the net phase acquired by a wavepacket in traversing the closed Feynman trajectory $\bar{\frako}_j$ illustrated in \fig{fig:interband}(a). ${\Omega}_{\bar{j}}{:}{=}\Omega[\bar{\frako}_j]$ has the same functional form as in \q{quantizationnecking}, with $e^{i\lambda}{=}e^{i\phi_{\sma{B}}}{\in}\R$ owing to $T\rot_{2z}$ symmetry; $e^{i\phi_{\sma{B}}}$ changes  discontinuously across the band touching point, owing to $\bar{\frako}_1$ encircling the Dirac point only for positive energies.
The opposing orientations of $\{\bar{\frako}_j\}$ result in $\{S[\bar{\frako}_j]\}$ carrying different signs.


Eqs.\ (\ref{quantizationIIdirac}) and (\ref{definenujberryonly}) is our central result for interband breakdown, and may be derived from eigen-solution of the effective Hamiltonian $\calh{=}(u{+}v\sz)K_x{+}w\sx K_y$ that is valid in the vicinity of the band-touching point, with kinetic quasimomentum operators satisfying $\bK{\times} \bK{=}i|e|\bB/c$; $\calh$ is written in a representation whose basis functions are magnetic analogs\cite{slutskin} of Luttinger-Kohn functions.\cite{Luttinger_Kohn_function} From $\calh$ we derive an improved connection formula\cite{AALG} which extends a previous work\cite{slutskin} by including the effect of the Berry phase; continuity of the connected WKB wavefunction then imposes Eqs.\ (\ref{quantizationIIdirac}) and (\ref{definenujberryonly}). 


Since no symmetry in any (magnetic) space group relates an electron to a hole pocket, the two harmonics $({\Omega}_{\bar{1}}{\pm}{\Omega}_{\bar{2}})/2$ in  \q{quantizationIIdirac} are generically incommensurate, and competitively produce a quasirandom Landau/dHvA spectrum. There are two semiclassical limits where a locally-periodic spectrum emerges: (i) for $\bar{\mu}{\gg} 1$, which is the weak-field limit above or below the Dirac-point energy, the intraband-transmission amplitude $\tau e^{i\bar{\varphi}}{\rightarrow}1$, and we obtain independent quantization conditions  ${\Omega}_{\bar{j}}{=}2n\pi$ for two uncoupled orbits. (ii)  For $\bar{\mu}{\approx}0$, the interband-tunneling probability approaches unity, and \q{quantizationIIdirac} is solved approximately by 
\e{ {l^2\big(\;{S}_{\bar{1}}+S_{\bar{2}}\;\big)}\big|_{E_n^0}= 2n\pi; \as S_{\bar{j}}:={S}[\bar{\frako}_j], \as n\in \Z. \la{faninter}}
Note that the $E_0^0$ level is field-independent, and it lies where the zero-field electron and hole pockets are perfectly compensated ($|{S}_{\bar{1}}|{=}|{S}_{\bar{2}}|$); generically, this is not the energy of the Dirac point.

One subtlety of the limit $\bar{\mu}{\rightarrow}0$ is that ${\Omega}_{\bar{j}}$ is well-defined only for isolated orbits [cf.\ \q{definenujberryonly}]. At the energy of the Dirac point, the two orbits merge into a figure of eight illustrated in \fig{fig:interband}(b), and the Berry connection (for a $\bk$-derivative in the azimuthal direction) diverges at $\bkp{=}0$.\cite{zak_compositebands} 
The validity of  \q{faninter} at strictly-zero energy  may  independently be justified by the following semiclassical quantization rule: to leading order in $|\bB|$, \q{faninter} may be re-interpreted as a generalization of the Onsager-Lifshitz rule\cite{onsager,lifshitz_kosevich} to an  orbit which is only partially electron-like.\cite{obrien_breakdown,koshino_figureofeight} The field-independent correction to \q{faninter} comprises the Maslov ($\phi_{\sma{M}}$) and Berry ($\phi_{\sma{B}}$) phases, which \emph{individually} vanish; this contradicts a  claim\cite{obrien_breakdown}  that $\phi_{\sma{M}}{=}\phi_{\sma{B}}{=}\pi$. That $\phi_{\sma{B}}$ vanishes follows from a pseudospin argument given in \fig{fig:interband}(b): by following the pseudospin texture (blue arrows) in a figure-of-eight trajectory, one finds that the wavefunction does not wind. $\phi_{\sma{M}}$ may be derived as the net phase in the connection formulae  of all turning points, where the WKB wavefunction is invalid.\cite{zilberman_wkb} The connection phase at each point is $\pm \pi/2$, where the sign is determined by the orientation of a wavepacket as it rounds the point.\cite{AALG} As indicated by green dots in \fig{fig:interband}(b), the net phase of the four turning points on the figure-of-eight vanishes, hence $\phi_{\sma{M}}{=}0$.




We now develop a perturbative treatment of quasirandom Landau/dHvA spectra, which applies in parameter regimes where one harmonic is dominant over the other. For $\bar{\mu}{\approx}0$, the dominant harmonic is associated to the semiclassical Landau fan: $\{E_n^0(B)\}_{\sma{n\in \Z}}$ which solves \q{faninter}; to leading order in the tunneling parameter $\tau$, the quantum correction to the fan oscillates with the frequency of the weaker harmonic [$({\Omega}_{\bar{1}}-{\Omega}_{\bar{2}})/2{+}\bar{\varphi}$]:
\e{\delta E_n^1 =2(\mo)^{\sma{n+1}}\text{sign}[E]\tf{\tau(\bar{\mu})}{l^2(S_{\bar{1}}+S_{\bar{2}})'}\text{sin}\big[\tf{l^2(S_{\bar{1}}-S_{\bar{2}})}{2}+\bar{\varphi} \big],\la{quantumcorr}}
with the right-hand side evaluated at $E_n^0$, and  the shorthand $O'{:}{=}\partial O/\partial E$. In particular, the quantum correction to the zeroth Landau level is a sinuisoid whose amplitude is linear in $E_0^0$ and grows as $|\bB|^{\sma{1/2}}$ to lowest order in $|\bB|$. As $|E|{\rightarrow} 0$, there is a logarithmic divergence in the second-order derivatives of the classical action function $[l^2(S_{\bar{1}}{-}S_{\bar{2}})]$  with respect to $E$; in \q{quantumcorr}, this non-analyticity is cancelled by a logarithmic divergence in the tunneling phase $\bar{\varphi}$. While the Berry phase did not affect the semiclassical Landau fan of \q{faninter}, its effect on the quantum correction is to shift the phase of $\delta E_n^1$ by $\pi/2$; this has already been accounted for in \q{quantumcorr}.  The validity of \q{quantumcorr} relies on $\tau$ and $\bar{\varphi}$ being small and slowly varying on the scale of $\delta E_n^1$. Indeed, the typical scale of variation for $\tau$ and $\bar{\varphi}$ is $\Delta \bar{\mu} {\sim} 1$, which implies an energy scale $\Delta E {\sim}  {\sqrt{{w}}(u^2-v^2)^{3/4}}/(vl).$ For typical values of $u$ and $v$, $\delta E_n^1/\Delta E$ vanishes  for  small enough field or $|E_n^0|$. 

The validity of our perturbation theory [\qq{faninter}{quantumcorr}] is tested against the numerically-exact magnetic energy levels of an over-tilted Weyl point. These levels are obtained by large-scale diagonalization of  the Peierls-substituted tight-binding model in Ref.\ \onlinecite{obrien_breakdown}. Inserting their tight-binding parameters (as detailed in the Supplementary Information) into \qqq{faninter}{quantumcorr}, we plotted in \fig{fig:interband}(d) the semiclassical fan [red, dashed lines] and the quantum correction [black solid], which compares favorably  with Fig.\ 2 in Ref.\ \onlinecite{obrien_breakdown}.

\noindent \emph{Discussion} We have presented generalized quantization rules that incorporate both quantum tunneling and the geometric Berry phase. Due to the intrinsic phase ambiguity in the wavefunction of wavepackets that approach/leave a tunneling region, we broadly argue that the geometric phase should appear in any tunneling phenomena. This phase is especially relevant if tunneling occurs within a subspace of states (bands, in our context) nontrivially embedded in  a larger Hilbert space; this point has been overlooked in conventional treatments\cite{azbel_quasiclassical,slutskin,berry_mount_review} of tunneling by connection formulae. 

The modern prototype of a nontrivially-embedded band is one that touches another at a conically-dispersing wavevector (a Dirac-Weyl point). We have exemplified how the unremovable geometric phase of a Dirac-Weyl point influences the quantization rules for both intra- [cf.\ Eqs.\ (\ref{quantizationnecking}),(\ref{definenujberryonly})] and interband [cf.\ Eqs.\ (\ref{quantizationIIdirac}),(\ref{definenujberryonly})] breakdown; consequences have been discussed for the spectra of Landau levels and dHvA peaks.

\begin{acknowledgments}
The authors are grateful to T. O'Brien for clarifying his numerical calculation. We acknowledge support by  the Yale Postdoctoral Prize Fellowship and NSF DMR Grant No.\ 1603243.
\end{acknowledgments}

\appendix

\section{Tight-binding parameters of the O'Brien-Diez-Beenakker  model} \la{app:obrienparameters}

We extracted the following tight-binding parameters from visual inspection of Fig.\ 3 in Ref.\ \onlinecite{obrien_breakdown}:
\e{  E_0^0 \eq -0.02\left(1+\f{120}{152}\right);\lin 
 a_0^2 S_1'(E_0^0) \eq \f{1.5+2.5/13}{0.2+0.05*50/76}; \lin
a_0^2 S_2'(E_0^0) \eq -\f{1.5+4/13}{0.3}; \lin
a_0^2 S_1(E)  \eq 1 + \f1{2}\f{94}{130} + E a_0^2S_1'(E_0^0),\lin
a_0^2 S_2(E)  \eq \f1{2}\f{23}{13} + E a_0^2S_2'(E_0^0),\lin
\bar{\mu}(E,B) \eq \f{0.52}{2\pi}\f{E^2}{B ea_0^2/h},}
with $S'{:}{=}\partial S/\partial E$, the units $c=1$ (speed of light), $t=1$ (a tight-binding hopping parameter); $a_0$ is a lattice constant that is assumed small relative to the magnetic length $l$, but is otherwise arbitrary.

\bibliography{bib_June2017}

\begin{thebibliography}{64}
\bibitem [{AAL()}]{AALG}%
  \BibitemOpen
  \href@noop {} {}\bibinfo {note} {The first-order-corrected connection
  formulae for both intra- and interband breakdown is presented in a separate,
  mathematically-oriented work (A. Alexandradinata and L. Glazman, in
  preparation).}\BibitemShut {Stop}%
\end{thebibliography}

\end{document}